\documentclass[%
 aip,
 amsmath,amssymb,
 reprint,%
]{revtex4-1}

\usepackage{graphicx}% Include figure files
\usepackage{physics}        % Dirac Notes
\usepackage{subfigure}      % combine Figures
\usepackage{dcolumn}% Align table columns on decimal point
\usepackage{bm}% bold math
\usepackage[utf8]{inputenc}
\usepackage[T1]{fontenc}
\usepackage{mathptmx}
\usepackage{nameref,hyperref}

\begin{document}

\preprint{AIP/123-QED}

\title{Numerical Study of Evaporative Cooling in the Space Station}
% Force line breaks with \\

\author{Bo Fan}
\email{bo.fan@pku.edu.cn}
 \affiliation{School of Electronics Engineering and Computer Science, Peking University, Beijing 100871, People's Republic of China}%Lines break automatically or can be forced with \\
 
\author{Luheng Zhao}%
%  \email{Second.Author@institution.edu.}
\affiliation{ 
School of Physics, Peking University, Beijing 100871, People's Republic of China%\\This line break forced with \textbackslash\textbackslash
}%

\author{Yin Zhang}
 \affiliation{School of Electronics Engineering and Computer Science, Peking University, Beijing 100871, People's Republic of China}
 
\author{Jingxin Sun}
\affiliation{School of Electronics Engineering and Computer Science, Peking University, Beijing 100871, People's Republic of China%\\This line break forced% with \\
}%
\author{Wei Xiong}
\affiliation{School of Electronics Engineering and Computer Science, Peking University, Beijing 100871, People's Republic of China%\\This line break forced% with \\
}%

\author{Jinqiang Chen}
\affiliation{Technology and Engineering Center for Space Utilization, Chinese Academy of Sciences, Beijing, China%\\This line break forced% with \\
}%

\author{Xuzong Chen}
\email{xuzongchen@pku.edu.cn}
\affiliation{School of Electronics Engineering and Computer Science, Peking University, Beijing 100871, People's Republic of China%\\This line break forced% with \\
}%

%------------------------------------------------------------------------------------
\date{\today}% It is always \today, today,

\begin{abstract}
In this paper, we numerically studied the effects of mechanical vibration and magnetic fields on evaporative cooling process carried in space station by direct simulation Monte Carlo method. Simulated with the vibration data of international space station, we found that the cooling process would suffer great atomic losses until the accelerations reduced tenfold at least. In addition, if we enlarge the s-wave scattering length five times by Feshbach resonance, the PSD increased to 50 compared to 3 of no magnetic fields situation after 5 seconds evaporative cooling. We also simulated the two stages crossed beam evaporative cooling process (TSCBC) under both physical impacts and obtain  $4\times10^5$ $^{85}$Rb atoms with a temperature of 8 pK. These results are of significance to the cold atom experiments carried out on space station in the future. 
\end{abstract}

\maketitle
%%%%%%%%%%%%%%%%%%%%%%%%%%  body  %%%%%%%%%%%%%%%%%%%%%%%%%%
\section{Introduction}
Pursuing further low temperatures has always been an important goal of ultracold atomic physics. Due to the influence of gravity, the conventional evaporative cooling process \cite{ketterle1996evaporative} cannot obtain a lower temperature by infinitely reducing the frequency of potential well. Generally, the temperature of order nanokelvin scale can be obtained \cite{jaksch1998cold}. Even if the gravity is balanced by magnetic levitation on the earth, it is difficult to further reduce the temperature significantly. For example, in 2003, Ketterle's group cooling atoms by evaporation and thermal diffusion under magnetic levitation conditions to obtain 450 pK temperature \cite{leanhardt2003cooling}, while in 2015, Kasevich's group obtained 45 pK transverse temperature by a transverse DKC cooling method \cite{kovachy2015matter}. In order to get the low temperature of pK or even fK level, researchers have chosen to carry out experiments in microgravity environment in recent years to overcome the influence of gravity. These experiments include tower drop experiments ($10^{-5}$g, QUANTUS, 2010, 2011) \cite{van2010bose,rudolph2011degenerate}, sounding rocket experiments ($10^{-5}$g, MAIUS 2015,2016) \cite{rudolph2015high, kubelka2016three}, parabolic flight experiments ($10^{-2}$g, ICE, 2010)\cite{geiger2011detecting}, NASA CAL team's experiments on the International Space Station($10^{-6}$g, 2016, 2018) \cite{lundblad2016progress,elliott2018nasa} and Chinese space station experiments scheduled for 2022. At the same time, a variety of deep cooling methods suitable for microgravity environments have emerged. The most typical two are the Delta Kick Cooling method (DKC) appeared in 1997 \cite{ammann1997delta} and the Two Stage Cross Beam Cooling method (TSCBC simplified to TSC) carried in all-optical trap \cite{wang2013generating, tian2014optimized, luan2015two, yao2016comparison}. The TSC method was proposed by the PKU research team in 2013 and verified by ground experiments at 2018 \cite{luan2018realization}. All these experiments and the emergence of new cooling mechanisms show the unanimous acceptance of cold atom experiments in microgravity environment. This paper takes the space station as an example to discuss how to obtain lower temperature in microgravity environment.

Besides many advantages such as good microgravity environment, unlimited experimental time, a spacious experimental space, there are still problems involving experiments in the space station. Mechanical vibration is the one must be taken into consideration. In general, vibrations have two effects mainly. First, the experimental devices may be damaged by severe vibration, which can always be avoided by engineering methods; Second, vibrations may interfere with the mechanism of cooling. Therefore, we use a numerical method to study the cooling process of rubidium atomic gas system under weak vibration during the operation phase of the space station. Using Direct Simulation Monte Carlo (DSMC) method \cite{bird1994molecular}, we consider the influence of the weak disturbance of the electromagnetic field caused by vibration on the mechanical motion of atoms and the change of potential well boundary. Based on the acceleration data \cite{everyspec.com} of the International Space Station (ISS), we simulated the evaporative cooling process and found that the vibration caused serious atomic number loss, especially in the low-frequency range. We will elaborate on this in the second section.

%ISS data 图
\begin{figure}[!htbp]
\centering
\includegraphics[width=0.5\textwidth]{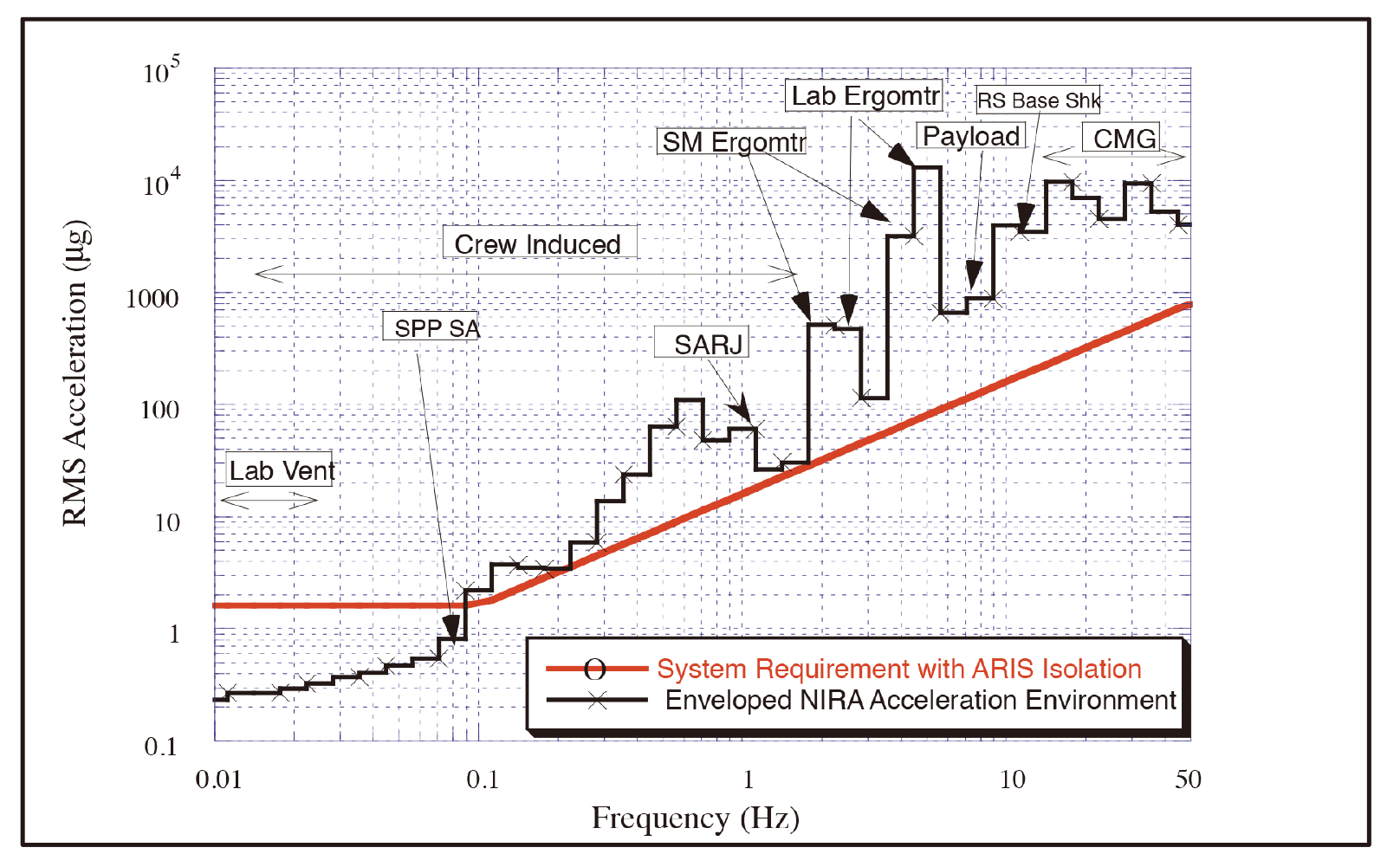} 
\caption{ Acceleration Data \cite{everyspec.com} of International Space Station(ISS)
\label{fig:ISS}}
\end{figure}

Shortening the degeneracy process by increasing phase space density (PSD) makes sense for ultra-cold physics experiments under vibration conditions. Through numerical simulation, we find that evaporative cooling with magnetic field regulation by Feshbach resonance \cite{chin2010feshbach} can greatly increase the PSD, thus achieving quantum degeneracy faster. In the third section, we will elaborate on the relationship between the external magnetic field, the scattering length of s-wave and the phase space density near Feshbach resonance. In the fourth section, we will simulate the TSC process under the influence of mechanical vibration of the space station and external magnetic field and finally obtained $4\times10^5$ degenerate $^{85}$Rb atoms with a temperature of 8 pK.

%--------------------------------------------------------------
\section{Simulation of Mechanical Vibrations Effects} % 第二部分

During normal operation phase of the space station, there are lot of mechanical vibration sources with different frequency. The working mechanical equipment, the flowing liquid and gas for environment controlling, and even flight attitude adjustment could result in weak vibration. We consulted micogravity environment data {\bfseries(Figure\ref{fig:ISS})} from NASA about the international space station(ISS) to simulate vibration influence. We simplified the problem and assumed that the vibration which atom `feel' is totally transmitted from space station cabinet where the cold atom experiment platform installed, and every parts of the platform works well. So that mechanical vibration cause only the trap shaking as a whole without any geometry distortion. Therefore, we believe that the movements of atoms in the optical dipole trap with external mechanical vibrating affection is a typical base exited forced vibration, comply with the equation as follows
% ------------ equations -----------
\begin{equation} \ddot{x}+\omega_0^2 x= A\omega_0^2\sin(\omega t + \phi_0) \end{equation}
where $x(t)$ represent the displacement of the particle, $\omega_0$ is for the frequency of trap, and  the external vibration frequency depicted by $\omega$. During time interval $\text{d}t $, the displacement of the particle and corresponding velocity will be

\begin{equation} 
\begin{split}
x(t) + \text{d}x &= - \frac{A\omega_0^2}{\omega^2-\omega_0^2}\sin(\omega \text{d}t + \phi_0) + \frac{v_0}{\omega_0} \sin{\omega_0 \text{d}t}\\ &+ x_0 \cos{\omega_0 \text{d}t} + \frac{A\omega_0}{\omega^2-\omega_0^2}[\omega_0 \cos{\omega_0 \text{d}t}+\omega \sin{\omega_0 \text{d}t}]
\end{split}
\end{equation}

\begin{equation} 
\begin{split}
v(t) + \text{d}v &= - \frac{A\omega_0^2 \omega }{\omega^2-\omega_0^2}\cos(\omega \text{d}t + \phi_0) + v_0 \cos{\omega_0 \text{d}t}\\ & - x_0\omega_0 \sin{\omega_0 \text{d}t} + \frac{A\omega_0^2}{\omega^2-\omega_0^2}[\omega \cos{\omega_0 \text{d}t} - \omega_0 \sin{\omega_0 \text{d}t}]
\end{split}
\end{equation}
% where K and G are
% \begin{align}
%   K = \left(x - \dfrac{A\omega_0^2}{\omega_0^2-\omega^2}\sin\omega t\right)\sin\omega_0t + \left(\dfrac{v}{\omega_0} + \dfrac{A\omega_0\omega}{\omega_0^2 - \omega^2}\cos\omega t\right)\cos\omega_0 t\\
%   G = \left(x - \dfrac{A\omega_0^2}{\omega_0^2-\omega^2}\cos\omega t\right)\sin\omega_0t + \left(\dfrac{v}{\omega_0} + \dfrac{A\omega_0\omega}{\omega_0^2 - \omega^2}\cos\omega t\right)\sin\omega_0 t
% \end{align}
% ---------------------

The mechanical vibration affected cooling process could be described also by quantum perturbation theory. The Hamiltonian is 

\begin{equation} 
H = \frac{P^2}{2m} + \frac{1}{2}m\omega_x^2[x - \epsilon_x(t)]^2
\end{equation}
where $\epsilon_x(t)$ is fluctuation of the trap center. The theoretical calculation of loading atoms in a shaking trap with constant trap depth $U_0$ given a result that the mean energy of atoms approaches a constant value 0.36 $U_0$ \cite{gehm1998dynamics} when system become stable. We simulate a similar heating process in which $4.7\times 10^4$ $^{87}$Rb atoms at 4.5 $\mu$K loaded in a crossed beam ODT with 1100 $\mu$K depth under 882 Hz external mechanical vibration. The final temperature become 340 $\mu$K and equal to 0.31 $U_0$. These have verified our simulation model.

%参量激发对比图
\begin{figure}[!htbp]
\centering
\includegraphics[width=0.45\textwidth]{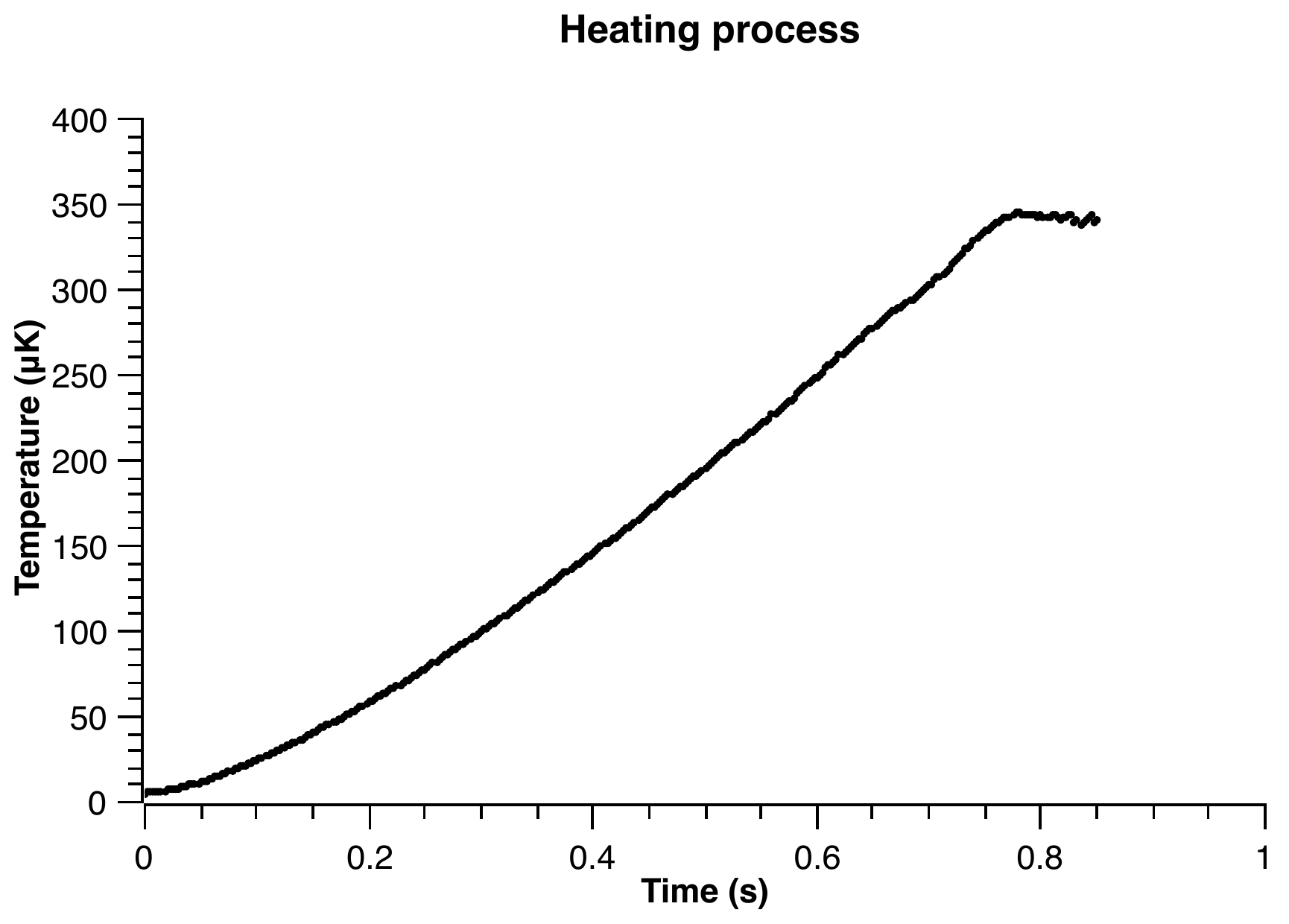} 
\caption{ $4.7\times 10^4$ $^{87}$Rb atoms at 4.5 $\mu$K heating by 882 Hz mechanical vibration in a crossed beam ODT with 1100 $\mu$K trap depth
\label{fig:Heating}}
\end{figure}

In the numerical simulation of mechanical vibration, we scan the frequency according to acceleration spectrum of ISS and study the responses of 5 seconds evaporative cooling process to vibrations. At the beginning, $7.5\times10^5$  $^{87}$Rb atoms with 10 $\mu$K temperature were loaded in an optical dipole trap (ODT) built by two crossed beams with 60 $\mu$m waist and 1064 nm wavelength both\cite{luan2018realization}. During the runaway evaporative process, the power of the beam is gradually reduced from 10 watts in 5 seconds. Power changed according to a nonlinear curve:
\begin{equation} 
P(t)= P_0\times \left(1+ \dfrac{t}{\tau}\right) ^\beta 
\end{equation}

%第一 & 第二幅图 %
\begin{figure*}[!htbp]
\centering

  \begin{minipage}{0.45\linewidth}
   \centerline{\includegraphics[width=1\textwidth]{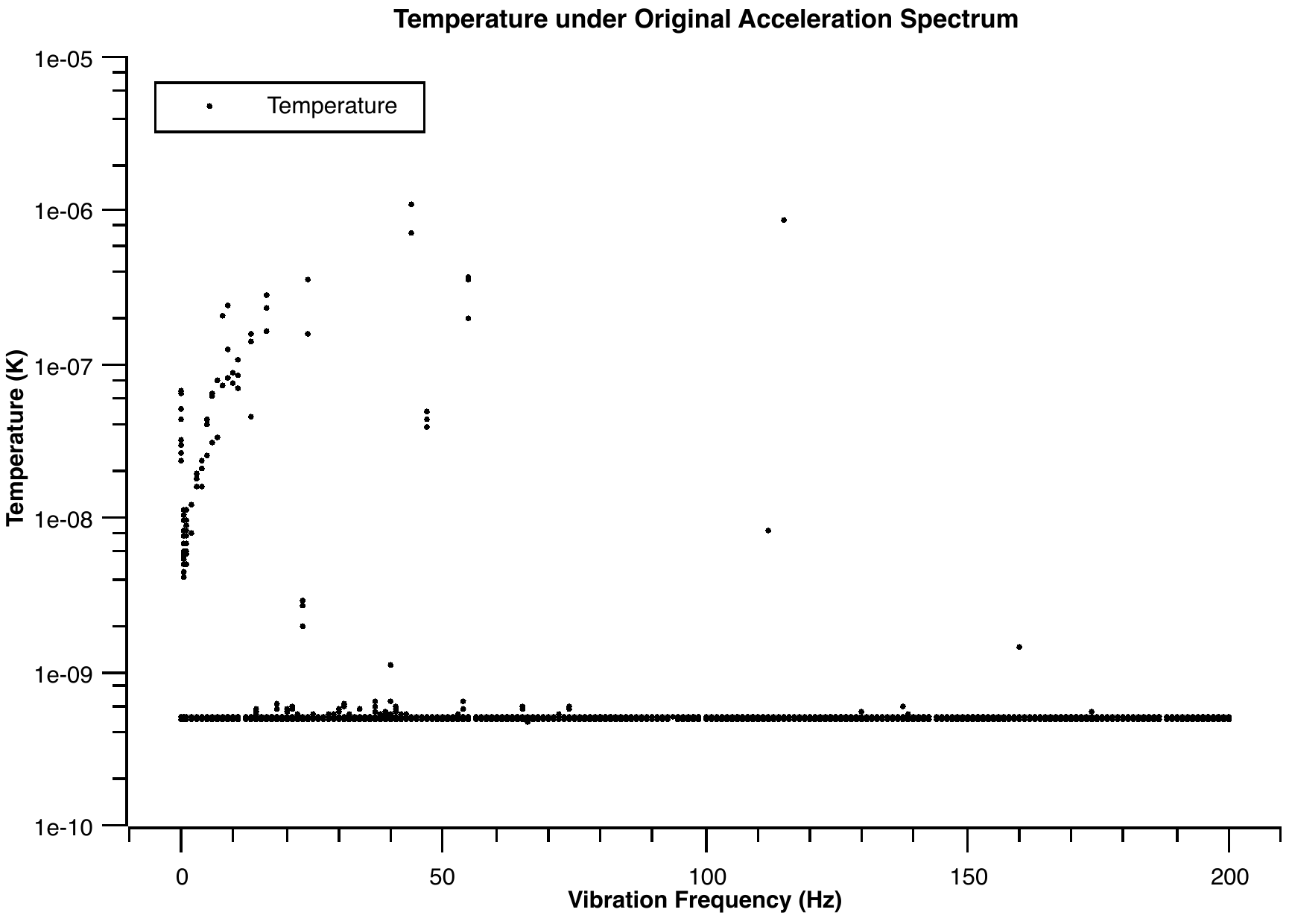}}
   \centerline{(a) final temperatures }
  \end{minipage}
   \qquad
  \begin{minipage}{0.45\linewidth}
   \centerline{\includegraphics[width=1\textwidth]{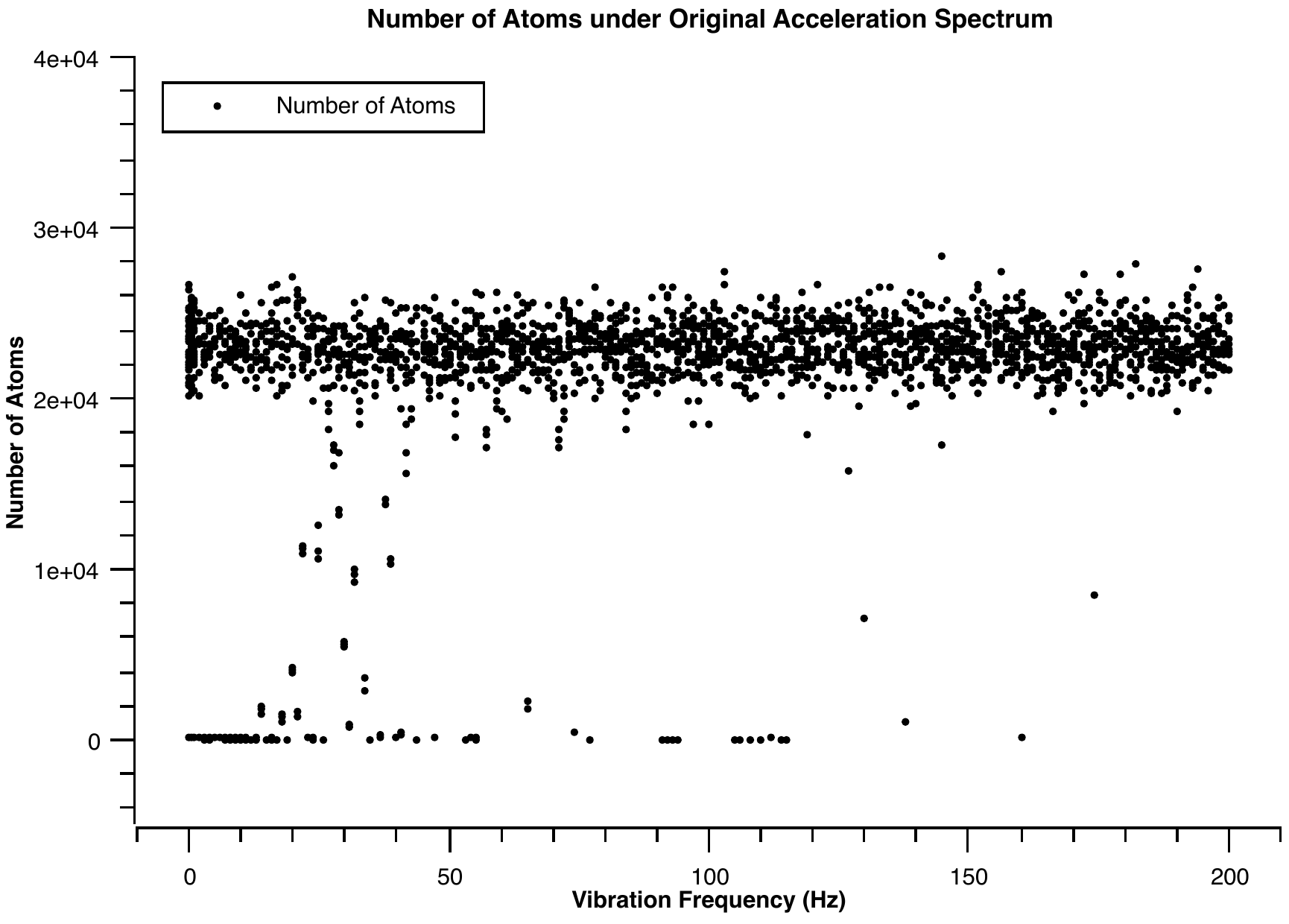}}
   \centerline{(b) final Number of Atoms }
  \end{minipage}

\caption{Vibrative affection analysis according to the original acceleration spectrum of international space station: (a) Temperature and (b) Number of $^{87}$Rb atoms after 5 Seconds evaporative cooling.   \label{fig:N_1xVib}} 
\end{figure*}
%----------------------------------------------------
\ref{tabel_1}%
\begin{table}[!htbp]
\caption{\label{tabel_1}Probability (Pr) of Serious Atomic Losses}
\begin{ruledtabular}
\begin{tabular}{llll}
    Freq (Hz)    \qquad \qquad & Pr(\%) @1X\footnotemark[1] \qquad& Pr(\%) @0.1X\footnotemark[2] \qquad & Pr(\%) @0.01X\footnotemark[3]  \\
\hline
    $0.1\thicksim0.9$    & $27$    & $29$  & $22$ \\  
    $1\thicksim10$       & $27$    & $21$  & $7$  \\   
    $11\thicksim20$      & $27$    & $6$   & $2$  \\  
    $21\thicksim30$      & $27$    & $0$   & $0$  \\  
    $31\thicksim40$      & $25$    & $0$   & $0$  \\ 
    $41\thicksim50$      & $10$    & $0$   & $0$  \\ 
    $51\thicksim60$      & $11$    & $2$   & $0$  \\ 
    $61\thicksim70$      & $0$     & $1$   & $0$  \\  
    $71\thicksim80$      & $6$     & $1$   & $0$  \\  
    $81\thicksim90$      & $0$     & $1$   & $0$  \\ 
    $91\thicksim100$     & $7$     & $1$   & $0$  \\  
    $101\thicksim200$    & $<2$    & $0$   & $0$  \\
\end{tabular}
\end{ruledtabular}
\footnotetext[1]{`1X' for original accelerations of ISS data.}
\footnotetext[2]{`0.1X' for one tenth.}
\footnotetext[3]{`0.01X for one per cent.}
\end{table}
%--------------------------------------------------------
Two coefficients equal to $\tau = 0.03$, $\beta = 2.78$. $P_0$ represent the initial power, and the corresponding trap frequency is $\omega_{x,y,z} = 2\pi\times(883,883,1248)$ Hz. Z axis is vertical to the crossed beam. According to our intuition,  mechanical vibrations work as a source of energy inputted in the cold atoms system will heat the atoms and get higher temperature. But the simulation results {\bfseries(Figure\ref{fig:N_1xVib})} show that there are hardly influence from the vibrations on the temperature of the system at the end of cooling if there still exist massive atoms in the trap. As an open system, the cold atom gas lost atoms and energy from time to time during the evaporative cooling. Therefore, the additional energy from vibration which transmitted into the optical trap is also lost, and the system temperature mainly depends on the variation of ODT depth during the cooling process.

%第三幅图 & 第四幅图
\begin{figure*}[!htb]
\centering
  \begin{minipage}{0.45\linewidth}
   \centerline{\includegraphics[width=1\textwidth]{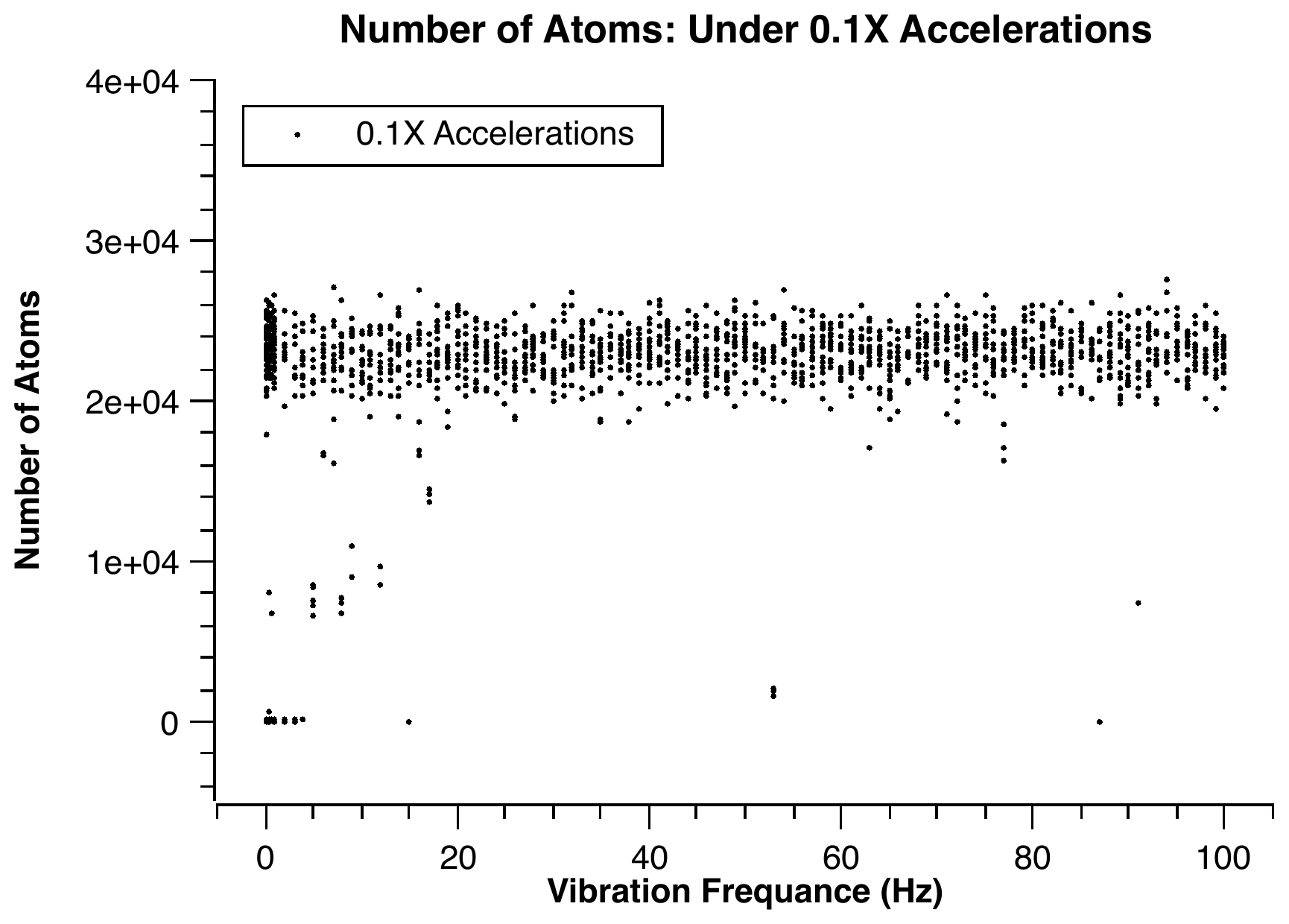}}
   \centerline{(a) 0.1X acceleration}
  \end{minipage}
  \qquad
  \begin{minipage}{0.45\linewidth}
   \centerline{\includegraphics[width=1\textwidth]{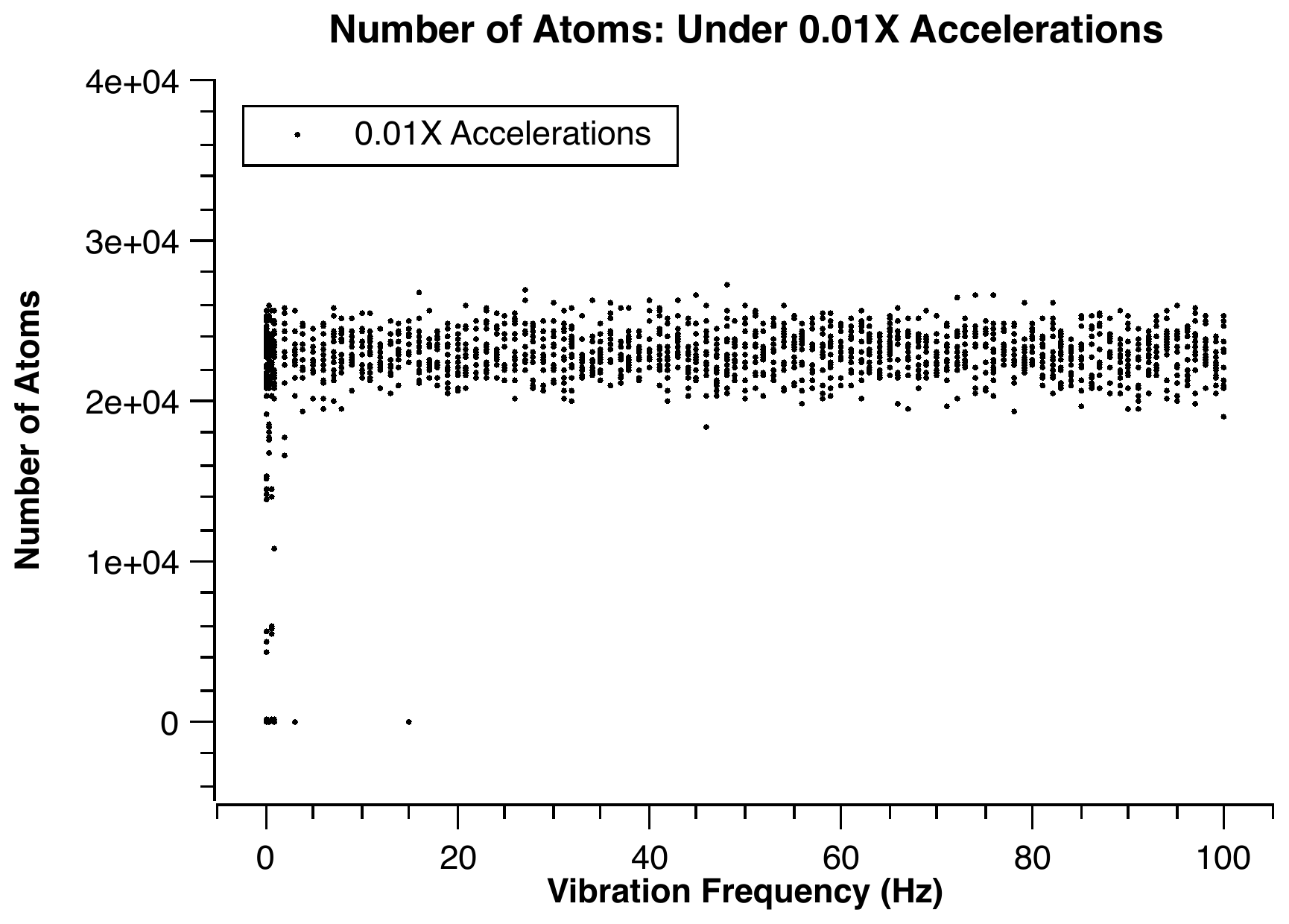}}
   \centerline{(b) 0.01X acceleration}
  \end{minipage}
\caption{Vibrative affection analysis according to the reduced acceleration spectrum of the international space station: number of atoms after 5 seconds evaporative cooling. (a)for one-tenth of original accelerations and (b) for one hundred times smaller accelerations
  \label{fig:N_0.1x_0.01XVib}} 
\end{figure*}
%-------------------------------------------------------------

Another obvious result is about the number of atoms after 5 seconds evaporative cooling. We simulated over 20 times in each single frequency from 0.1 Hz to 200 Hz with 1 Hz step (above 1Hz) in the acceleration spectrum. The vast majority of the cooling processes finally reached 50nK and the number of atoms {\bfseries(Figure\ref{fig:N_1xVib})} locate in the range from $2\times10^4$ to $2.8\times10^4$ with a standard deviation of $6\times10^3$. So we can define a `serious atom loss' when the final number less than $1.8\times10^4$. But in the low vibration frequency band, some processes got only thousands of atoms in the trap and even less. There are $27\%$ cooling process suffered serious atom losses which frequency under 50Hz and less $2\%$ in the range of 100 Hz to 200 Hz{\bfseries( Table\ref{tabel_1} )}. In the lower frequency band, the simulation shows that mechanical vibrations `threw' out too many atoms from the trap as the heating. 

If we reduce the vibration acceleration below 50 Hz by 10 times, only $6.5\%$ of the cooling process will suffer serious atomic number loss{\bfseries( Figure\ref{fig:N_0.1x_0.01XVib}(a) )}; if we reduce the acceleration by 100 times, $3.6\%$ of the cooling process will suffer serious atomic number loss {\bfseries( Figure\ref{fig:N_0.1x_0.01XVib}(b) )}. In this low-frequency range, the optical dipole trap has a large vibration displacement relative to the size of the optical trap. If there are significant amounts of hot atoms near the boundary of ODT trap, it will be a large opportunity for serious atom loss, especially in the late period of the cooling where optical dipole trap maintain very low trapping frequency. Conversely, in the high-frequency band, the vibratory amplitude is smaller, so that the probability of atom losses become lower. Facts that the suppression of acceleration results in lower atom losses indicates that great affords should be taken to reduce the acceleration by tenfold at least in the range 1 Hz to 100Hz. Detailed simulation results of atomic quantity can be found in the Table\ref{tabel_1}.

%------------------------------------
%------------------------------------
\section{Simulation of Magnetic tuned Evaporative Cooling Process}
With the development of optical lattice technology, ultra-cold atomic system have become convenient and powerful quantum simulators nowadays. Therefore, it is more important to achieve high quantum degeneracy in addition to lower the temperature. As a direct reflection of the quantum degeneracy process, increasing the phase space density (PSD) as soon as possible will shorten the process to obtain BEC, and is also conducive to subsequent experiments such as quantum phase transition. Here we first briefly analyze the mechanism by which the magnetic field regulates the inter-atomic scattering length to increase the PSD and then show the simulation results using the DSMC method.

\subsection{Brief analyses of the mechanism of Magnetic Tuning}
It can be seen from the relation $\rho \thicksim N\omega_x\omega_y\omega_z T^{-3}$ that the phase space density $\rho$ of the atomic system in an optical trap is determined by the trapping frequency, the number of atoms and the temperature. Maintaining a higher number of atoms and lowering the final temperature of a cooling sequence (keeping the same $\omega_{xyz}$ variation process) is good for increasing the PSD value.

Specifically, the mechanism of evaporative cooling is due to losing high energy atoms from trap to lower the system temperature. The lost energy during time interval $\Delta t$ can be expressed as $ \Delta E_{evap} = - N_L \overline{E} \Delta t$, where $N_L$ represent the number of lost atoms and $\overline{E}$ denotes the average energy per atoms. Thus, if a small number of high-energy atoms escaped, the atomic system could simultaneously lower the temperature, maintain a higher number of atoms and finally obtain a larger PSD value.

Simply put, a higher elastic scattering rate can achieve conditions above which help to obtain higher PSD. We know that the cooling process is not in equilibrium and the energy distribution of the atoms deviates from the Maxwell distribution law. During the cooling process, more atoms are distributed in a relatively high energy range. With the escaping of these large amounts of energetic atoms, the system will be cooled but obtained a relatively small PSD. Therefore, we could increase the rate of elastic scattering between atoms in evaporative cooling to reduce the relaxation time, so that the atomic distribution is closer to the Maxwell-Boltzmann distribution. At this point, the escape of these relatively few high energy atoms will increase the PSD and effectively reduce the system temperature.

For cold and ultra-cold atomic gas systems, the interaction between particles is depicted as low-energy s-wave scattering mainly. From the relation $\Gamma = 8\pi a^2$, where $\Gamma$ denotes the elastic scattering rate and $a$ for s-wave scattering length, we can easily notice that the larger scattering length the more elastic collisions in one time of collision. So, if we precisely tune $a$, it is possible to accelerate degenerate process. Near a Feshbach resonance, the s-wave scattering length between atoms can be manipulated by an external magnetic field B, and the relationship between them is as follows \cite{moerdijk1995resonances}:
 \begin{equation} a(B) = a_{bg}\left( 1- \dfrac{\Delta}{B - B_0} \right) \end{equation}

 The background scattering length, which is denoted as $a_{bg}$, lies far from resonance center in frequency spectrum and weakly dependent on the external magnetic field. `$\Delta$' represent the Feshbach resonance width and $B_0$ denotes the resonance position.
 %-----------------------------------------------------------
\subsection{ Results of PSD Increment by Magnetic Tuning }
There are many resonances for certain kind of atom gas in a large magnetic field range and it is important to select a suitable Feshbach resonance to carry out experiments. In addition to the scattering properties of atoms, the technical difficulties in experiments also need to be carefully considered. For example, a higher magnetic field requires higher current for coils, while a narrow resonance width harm to a fine adjustment of the s-wave scattering length. 

For $^{87}$Rb, there are more than 40 resonances in the range of 0.5G to 1260G \cite{marte2002feshbach}. The broadest resonance $(\Delta = 0.17G)$ located at 1007G for state $\ket{1,1}$. It is very high magnetic field but narrow width. The three-body inelastic collision rates $K_3$ here is quite larger $(K_3\thicksim 10^{-27}cm^6/s)$ \cite{marte2002feshbach}, compared to zero magnetic field case $(K_3 = 8.3\times10^{-30} cm^6 /s)$. It is quite difficult for experimental implementation. But magnetic tuning of interactions between $^{85}$Rb atoms are much convenient because of a weaker magnetic field for resonance with a large resonance width at 155Gauss \cite{cornish2000stable}. Usually, a lower magnetic field requires a lower magnetic field noise caused by the current. At the same time, a wider resonance $(\Delta = 10.7G)$ for $^{85}$Rb atoms at hyper-fine state $\ket{2 , -2}$ needs an easier current control\cite{cornish2000stable, claussen2003very, roberts1998resonant}. 

 %第五幅图 & 第六幅图
\begin{figure*}[!htbp]
\centering

  \begin{minipage}{0.45\linewidth}
   \centerline{\includegraphics[width=1\textwidth]{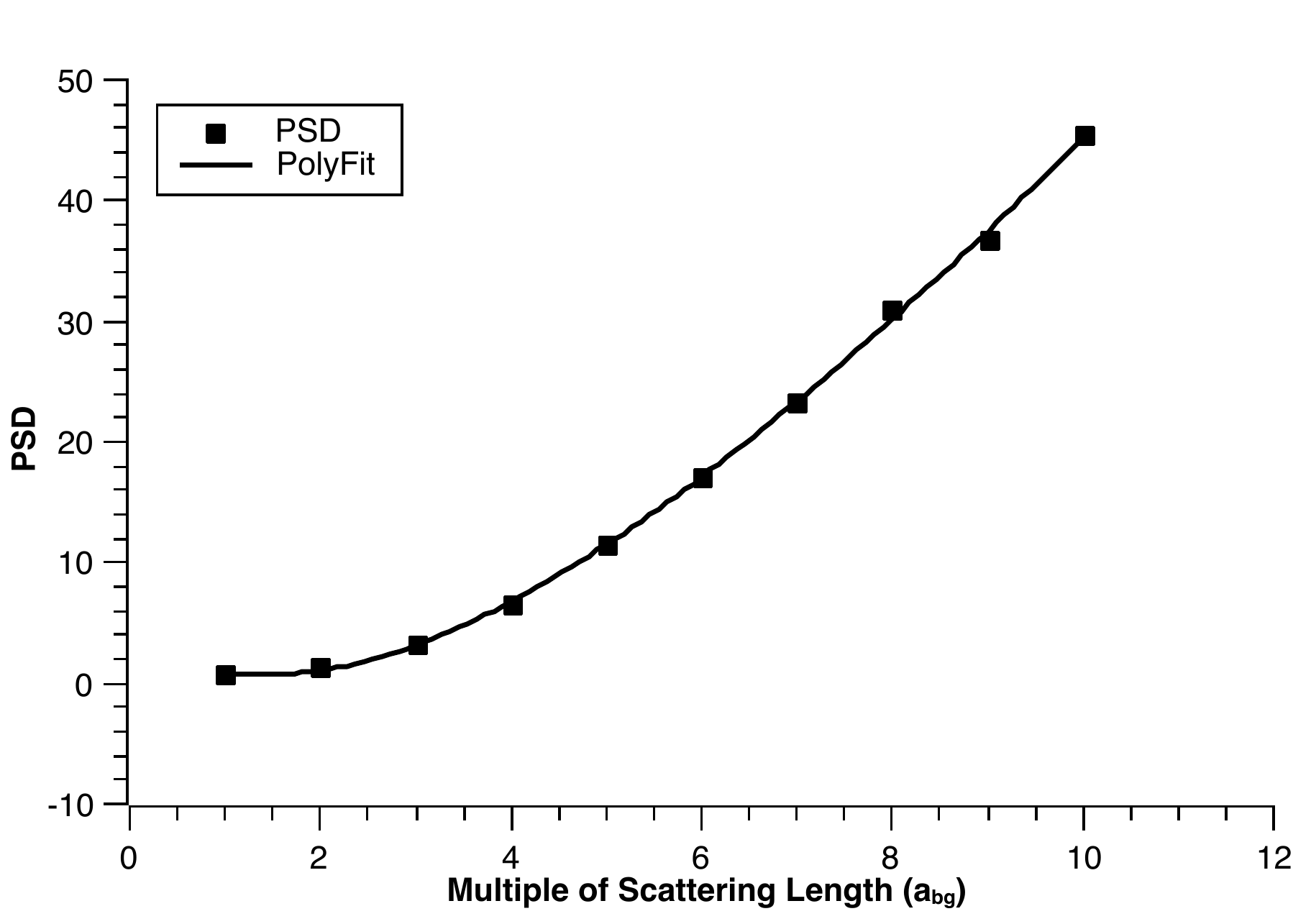}}
   \centerline{(a) phase space density}
  \end{minipage}
  \qquad
  \begin{minipage}{0.45\linewidth}
   \centerline{\includegraphics[width=1\textwidth]{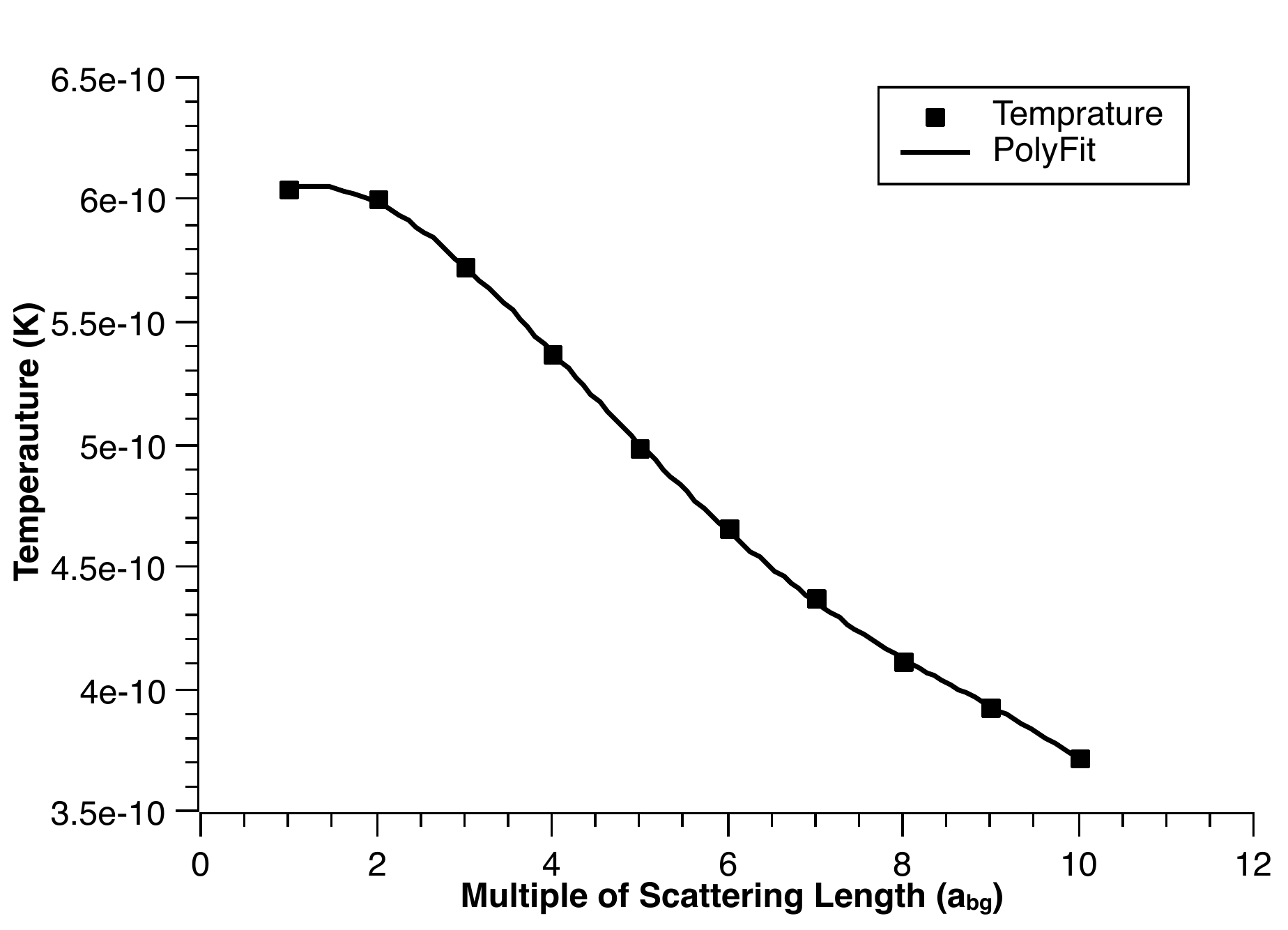}}
   \centerline{(b) temptrature}
  \end{minipage}

\caption{Analysis of magnetic tuning affection: responses of (a)phase space density and (b)temptrature to the varied s-wave scattering lengths.
  \label{fig:MagTuningPSD_Temp}} 
\end{figure*}

 Similar to the study of effects of vibration above, we simulated the magnetically tuned evaporative cooling process of $^{87}$Rb atomic gas for 5 seconds. The optical dipole trap setting, the laser power variation curve, the atomic quantity and the system temperature are the same as those of the aforementioned vibration studies.

\begin{figure}[!htb]
\centering
\includegraphics[width=0.45\textwidth]{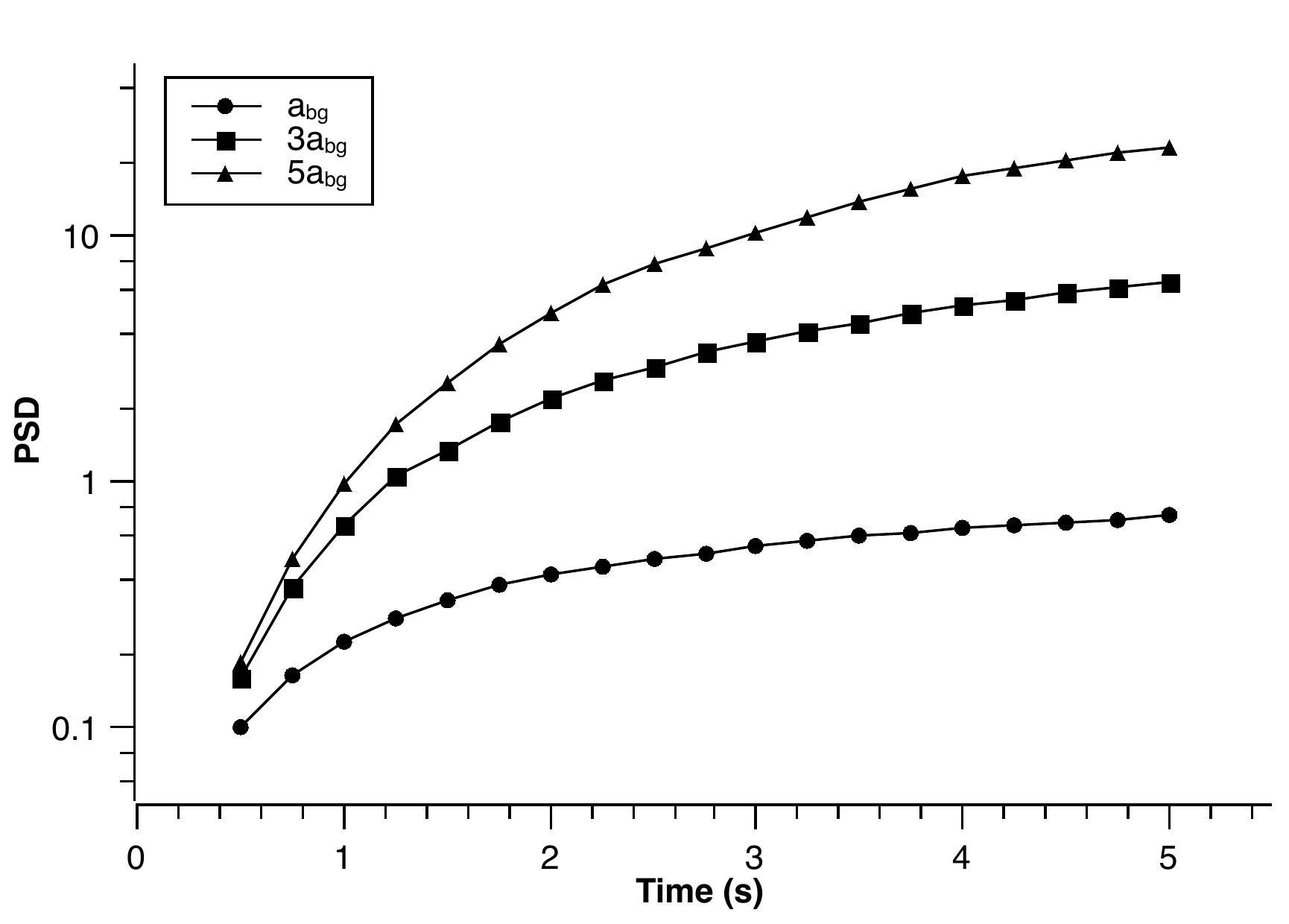} 
\caption{Analysis of magnetic tuning affection: responses of phase space density to three different scattering lengths ($a_{bg}, 3a_{bg}, 5a_{bg}$)
\label{fig:PSD_3Curve}} 
\end{figure}

 Generally speaking, the simulation results demonstrate that the increased s-wave scattering lengths due to magnetic tuning will increase the number of atoms in optical dipole trap {\bfseries( Table\ref{Table_2} )} and result in larger phase space densities {\bfseries( Table\ref{Table_2}, Figure\ref{fig:MagTuningPSD_Temp}(a) )}, but a slightly different temperatures {\bfseries ( Table\ref{Table_2}, Figure\ref{fig:MagTuningPSD_Temp}(b) )} at the end of evaporative cooling. Without any magnetic bias, scattering length equal to $a(0)\approx a_{bg}=100.44 a_0 $ \cite{harber2002effect}, the 5 seconds cooling eventually get $2.6\times10^3$ atoms at 60.4 nK without any condensate {\bfseries( Figure\ref{fig:PSD_3Curve} )}. Merely tripling scattering length, in which the system degenerated, got about $1\times10^5$ atoms condensate and PSD up to 3.1 compared with 0.7 without magnetic fields. The simulations also show that if we extend the scattering length extremely to 10 times that of the non-magnetic field, the temperature is slightly reduced but the PSD reaches 45.4 {\bfseries( Table\ref{Table_2} )}, which is nearly 60 times expanded. Of course, with such a large scattering length, the condensate is very unstable.
 
 %第七幅图

%---------------------------------------------
 %---------------------------------------------
\begin{table}[!hbpt]
  \caption{Simulation of Magnetic tuned Evaporative Cooling Process.}
  \begin{ruledtabular}
    \begin{tabular}{llll}
    Scattering Length   & Temperature (pK) & Number of Atoms & PSD \\
    \hline
    $a_{bg}$       & $604$  & 2583   & 0.7 \\
    $2a_{bg}$      & $600$  & 4868   & 1.4 \\ 
    $3a_{bg}$      & $573$  & 9677   & 3.1 \\  
    $4a_{bg}$      & $537$  & 16337  & 6.5 \\ 
    $5a_{bg}$      & $499$  & 23607  & 11.6 \\
    $6a_{bg}$      & $466$  & 28166  & 17.0 \\ 
    $7a_{bg}$      & $437$  & 31781  & 23.3 \\ 
    $8a_{bg}$      & $411$  & 35169  & 31.0 \\ 
    $9a_{bg}$      & $393$  & 36302  & 36.7 \\ 
    $10a_{bg}$     & $372$  & 38292  & 45.4 \\ 
    % \midrule
    \end{tabular}%
  \end{ruledtabular}%---------------------
  \label{Table_2}%
\end{table}%
%--------------------------------------------- 

\section{Simulation of TSC process in Space Station}
Numerical studies show that the Two Stage Crossed Beam Cooling (TSC) method can be used to cool the $^{87}$Rb atomic system to an ultra-low temperature of 7pK in an ideal microgravity environment\cite{yao2016comparison}. The effectiveness of the TSC method is also experimentally confirmed on the ground with the magnetic levitation method \cite{luan2018realization}. In this paper, we numerically studied the TSC cooling process with $^{85}$Rb atoms for deep cooling considering the effects of the actual microgravity environments in space station and external magnetic fields. Specifically, we simulated three TSC cooling process with same time sequence of the $^{87}$Rb atomic system under three different cases.  First, `normal' cooling process without external magnetic field and vibration. Second, magnetic tuned process with scattering length $(a =7a_{bg})$ and no vibration. Third, cooling with vibration from 1 Hz to 2000 Hz (randomly appearing) and magnetic tuned scattering length$(a =7a_{bg})$.

The 15 second typical TSC cooling process under these three cases can be divided into two stage. The first stage belongs to a 5 seconds runaway evaporative cooling process and the second is a adiabatic diffusing cooling process in 10 seconds. Correspondingly, the tight-confining optical trap of the first stage are built by two orthogonal beams with 60 um width, 1064 nm wave length, and an initial power of 12 W. The wide-bonded optical trap of second stage consists of two orthogonal beams with 350 um waist and 650 mW initial power. Two nonlinear coefficients have different values for two process, which is  $\tau_1 = 0.03$ $\beta_1 = 2.36$, $\tau_2 = 0.01$ $\beta_2 = 1.19$. There are $7.5\times10^5$ $^{85}$Rb atoms with 10 $  \mu$K temperature after optical molasses phase.

 %第八幅图 & 第九幅图
\begin{figure*}[!htb]
\centering

  \begin{minipage}{0.45\linewidth}
   \centerline{\includegraphics[width=1\textwidth]{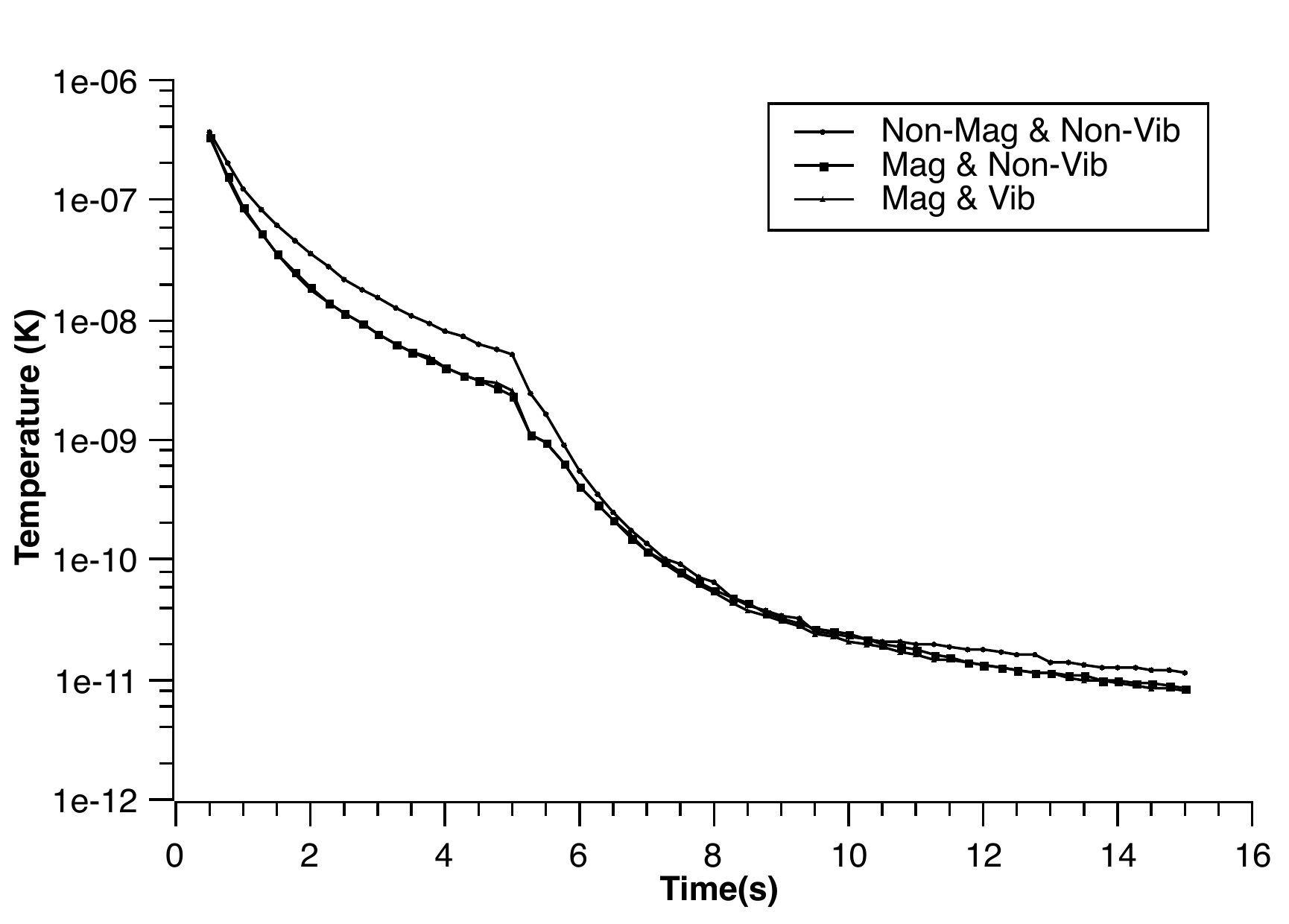}}
   \centerline{(a) Temperature}
  \end{minipage}
  \qquad
  \begin{minipage}{0.45\linewidth}
   \centerline{\includegraphics[width=1\textwidth]{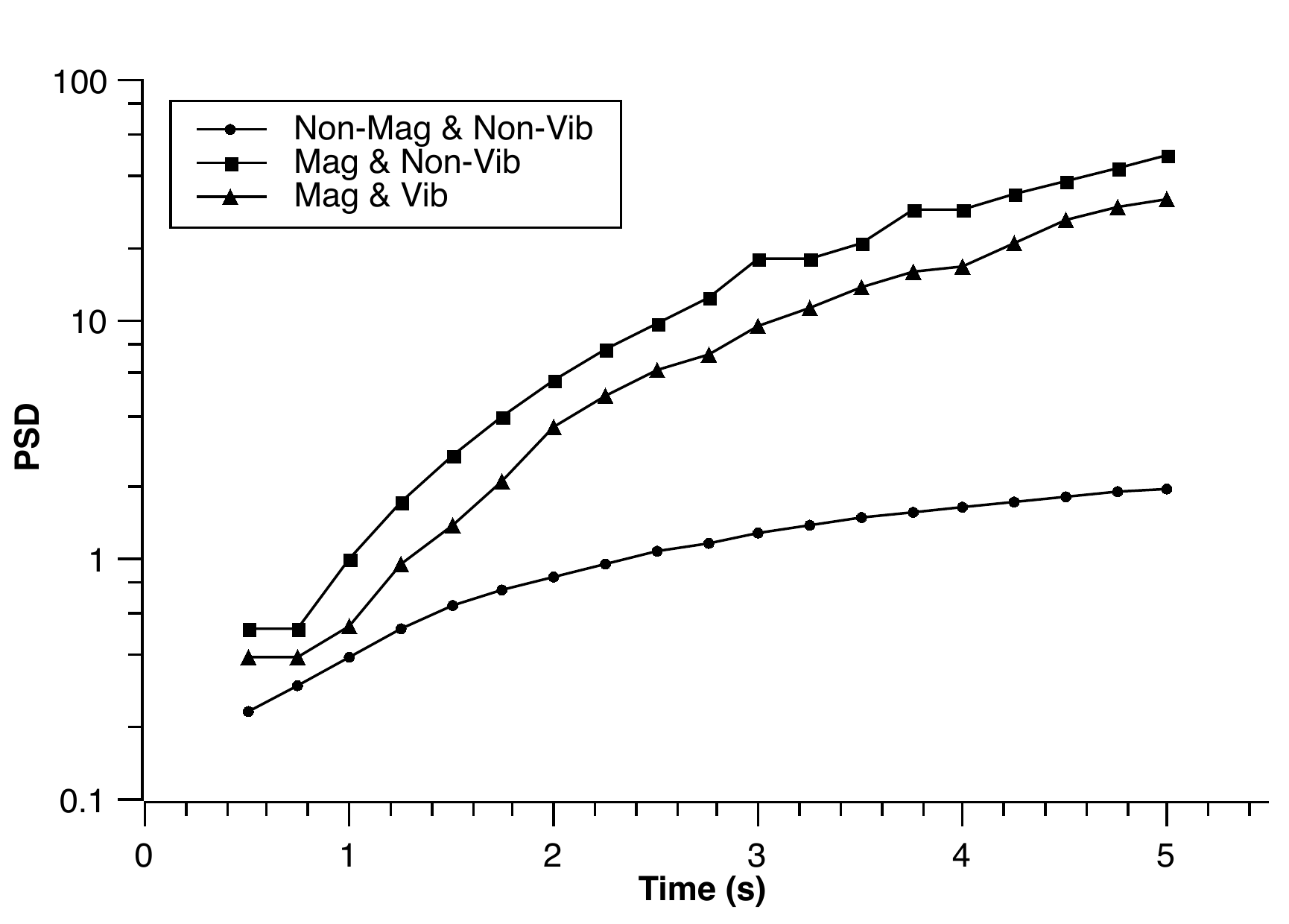}}
   \centerline{(b) Phase Space Density}
  \end{minipage}

\caption{Simulation of TSC Process under Three cases: 1.`normal' cooling process without external magnetic field and vibration 2.magnetic tuned process with scattering length and no vibration 3. cooling with randomly appearing vibration and magnetic tuned scattering length
  \label{fig:TSCBC}} 
\end{figure*}

The results are shown in {\bfseries( Fig.\ref{fig:TSCBC} )}, in which the temperatures are almost similar in three cases. The final temperatures achieve 12.3 $p$K for normal process, 8.6 $p$K for magnetic tuned process and 8.1 $p$K for the case of magnetic tuning and mechanical vibration affected{\bfseries( Figure\ref{fig:TSCBC} (a) )}. In the first stage of runaway evaporation cooling, magnetic tuned process with or without vibration have tiny difference, that is about $2.3n$K and $2.5n$K , compared with $5.2n$K in the normal process. These results would remind us again that temperature variation is dominated by trap depth mainly. 

Although the huge disparities of PSD curves {\bfseries( Figure\ref{fig:TSCBC} (b) )} could explain for the tiny difference about temperature, but the faster finishing of degeneration is of great importance. Without magnetic field, the PSD at end of first stage result in 1.9. In contrast, the other two processes with enlarged scattering length got higher values, which is 26.5 for vibration influenced one and 48.4 for the other. The vibrative affection results in a lower PSD than a vibration-free process due to large atom losses. This results also verify the necessity about the vibration isolation. We have to emphasize that the process under vibration conditions that we present here is done without substantial loss of atoms, which is occasionally happened.

\section{Conclusion}
In conclusion, with numerical method we study the cooling process of rubidium atomic gases in the environments of space station, including normal evaporative cooling and TSC deep cooling processes. The results showed that the enlarged scattering length could significant increasing phase space density of the cold atoms gas and shorten the process to get BEC in the crossed beam optical dipole trap. It is likely that this tuning method becomes a popular and powerful experimental tool for neutral atoms cooling acceleration. It is disappointing that the mechanical vibration can damage the cooling process by significant atom loss occasionally and the situation would very serious in the lower band of vibratory accelerations spectrum(ISS Data) which below 200Hz. Great measures should be taken to isolate vibrations and reduce the accelerations tenfold at least. So we know that applying vibration isolation and increasing scattering length by the magnetic tuning, one can carry out the ultra-cold atom experiments stably and faster in the space station environments, and finally reach the ultra-low temperature of about 8pK through the TSC process.

\begin{acknowledgments}
We thank Professor Shuyu Zhou for his helpful discussions and suggestions. This work is supported by the National Key Research and Development Program of China (Grant No. 2016YFA0301501), and the National Natural Science Foundation of China (Grants No. 91736208, 11334001, 61727819, 61475007).

This article has been submitted to [Review of Scientific Instruments]. After it is published, it will be found at \href{https://publishing.aip.org/resources/librarians/products/journals/}{Link}.
\end{acknowledgments}

\bibliography{Simu_Vib_MagneTun}
%%%%%%%%%%%%%%%%%%%%%%%%%%%%%    end      %%%%%%%%%%%%%%%%%%%%%%%%%%%%%%%

\end{document}